\documentclass[10pt,conference]{IEEEtran}
\usepackage{graphicx}

\begin{document}

\title{A Plotkin-Alamouti Superposition Coding Scheme for Cooperative Broadcasting in Wireless Networks}

\author{
\authorblockN{Robert Morelos-Zaragoza}
\authorblockA{Department of Electrical Engineering\\
San Jos\'e State University \\
San Jose, CA 95192-0084, USA\\
Email: r.morelos-zaragoza@ieee.org}
}
%

\maketitle

\begin{abstract}
This paper deals with superposition coding for cooperative broadcasting in the case of two coordinated source nodes, as introduced in the seminal work of Bergmans and Cover in 1974. A scheme is introduced for two classes of destination (or relay) nodes: {\em Close} nodes and {\em far} nodes, as ranked by their spatial distances to the pair of transmitting nodes. Two linear codes are combined using the $\vert u\vert u+v\vert$-construction devised by Plotkin to construct two-level linear unequal error protection (LUEP) codes. However, instead of binary addition of subcode codewords in the source encoder, here modulated subcode sequences are combined at the destination (or relay) nodes antennae. Bergmans and Cover referred to this as {\em over-the-air mixing}. In the case of Rayleigh fading, additional diversity order as well as robustness to channel estimation errors are obtained when source nodes transmit pairs of coded sequences in accordance to Alamouti's transmit diversity scheme. We refer to this combination as a Plotkin-Alamouti scheme and study its performance over AWGN and Rayleigh fading channels with a properly partitioned QPSK constellation.
\end{abstract}

\section{Introduction}
Most recently, the amount of research devoted to coope\-\-rative wireless networks has grown considerably~\cite{[3]}-\cite{[9]}. An important aspect of these networks is that communication between network nodes takes place in two stages: (1) broadcasting and (2) multiple access. In 1974, Bergmans and Cover~\cite{[2]} showed that coordinated sources broadcasting information using superposition coding~\cite{[1]} always outperform other schemes based on orthogonal channel assignments, such as time sharing or frequency division. Surprisingly, most of the recent work in this area omits reference to the seminal work in~\cite{[2]}. Therefore one goal of this paper is to bring attention to the fact that the fundamental idea of cooperative broadcasting was introduced thirty five years ago. 

In this paper, superposition coding by two cooperating source nodes that broadcast information to two types of destination/relay nodes is considered. This is the same set up as in reference~\cite{[2]}. To provide two levels of error protection, the $\vert u\vert u+v\vert$-construction devised by Plotkin~\cite{plotkin} is employed. Important is the fact that, contrary to the binary case in which addition is modulo two, superposition of {\em modulated} $\vert u\vert u\vert$ and $\vert 0\vert v\vert$ subcode signal sequences takes place at the receive antennae, with addition over the field of real numbers (or, depending on the signal constellation selected, over a multi-dimensional real-number field). To achieve two levels of error protection, a signal constellation needs to be partitioned into two sets in such a way that a demapper at the receiver can produce meaningful metrics for soft-decision decoding. In essence, this means that the signal constellation is partitioned into subsets forming a so-called Sidon set~\cite{uep_ldpc}. For the sake of simplicity of exposition, in this paper focus is on a symmetric QPSK constellation and an example of a two-level superposition code and its error performance are presented.

The rest of the paper is organized as follows: In section~\ref{sec2}, a superposition code construction for two classes of destination nodes will be introduced. This work is based on fundamental ideas put forward in the work of Cover~\cite{[1]} on broadcast channels. A specific example using Plotkin's construction will be introduced and shown to achieve two levels of error protection using two independently chosen constituent codes and a simple condition on their minimum Hamming distances. 

For transmission in a wireless network using a Rayleigh flat-fading channel model, superposition coding can be combined with Alamouti's transmit diversity scheme~\cite{alamouti}. This is the topic of section~\ref{sec3}. The main advantages derived from the use of Alamouti's transmission method are an increase in the diversity order and a robustness to channel estimation errors. Section~\ref{sec4} gives a specific example of the proposed superposition coding construction scheme, with short and simple constituent codes: A binary single-parity-check $(20,19,2)$ code and a binary Gallager $(20,7,6)$ LDPC code. The use of relatively short codes is supported by the assumption of a block fading channel needed in Alamouti's scheme. Error performance results of this superposition code over AWGN and Rayleigh channels with belief propagation iterative decoding are presented. Future research topics and concluding remarks are given in section~\ref{sec5}.

\section{Superposition coding}
\label{sec2}
The main purpose of this paper is to present a construc\-\-\-tive example of superposition coding for cooperative broadcasting with two classes of destination/relay nodes~\cite{[2]}: Close nodes and far nodes, as ranked by their spatial distance to a pair of coordinated source nodes. Those destination or relay nodes that are close to a source node are able to receive all of the transmitted information bits with a high degree of reliability, while nodes far from source nodes can only recover reliably a portion of the transmitted bits. Therefore, two levels of error protection are required.
To provide two levels of error protection, in this paper the $\vert u\vert u+v\vert$ construction~\cite{plotkin} will be used. Let $C_1$ and $C_2$ be two linear $(n,k_1,d_1)$ and $(n,k_2,d_2)$ codes, respectively. Then the linear code
\[ C \buildrel\Delta\over = \{ \vert \bar{v}_1\vert \bar{v}_1+\bar{v}_2\vert : \bar{v}_1\in C_1, \bar{v}_2\in C_2 \}, \]
is a linear $(2n,k_1+k_1,\min\{2d_1,d_2\}$ code with generator and parity-check matrices
\[ G = \pmatrix{G_1 & G_1\cr 0 & G_2}, \quad H = \pmatrix{H_1 & H_1\cr 0 & H_2}, \]
respectively. If the condition $2d_1 < d_2$ is satisfied~\cite{vangils}, then code $C$ is said to be a linear unequal-error-protection (LUEP) code with separation vector $\bar{s} = (d_2,2d_1)$ for the message space $M=\{0,1\}^{k_2}\times \{0,1\}^{k_1}$. This LUEP code is designed for a degraded binary broadcast channel. 

\subsection{Over-the-air mixing}

In the case of over-the-air mixing~\cite{[2]}, or superposition, the subcode codewords $\vert \bar{v}_1\vert \bar{v}_1\vert$ and $\vert \bar{0} \vert \bar{v}_2\vert$ are modulated and sent separately by each of the two source nodes. These signal sequences are then combined by the receive antennae at the destination (or relay) node. Let $m_1(\vert \bar{v}_1\vert \bar{v}_1\vert)$ and $m_2(\vert \bar{0} \vert \bar{v}_2\vert)$ denote the mappings of bits to signal sequences drawn from two  (generally two-dimensional) signal sets, ${\cal M}_1$ and ${\cal M}_2$, each associated to a source node. Then  received signal sequence is given by
\[ \bar{y} = m_1(\vert \bar{v}_1\vert \bar{v}_1\vert) + m_2(\vert \bar{0} \vert \bar{v}_2\vert)+\bar{n}, \]
where addition (per dimension) is over the field of real numbers, $m_i(\cdot)\in {\cal M}_i$, $i=1,2$, and $\bar{n}$ is a two-dimensional Gaussian random vector with independent zero-mean components of equal variance $N_0/2$.

\subsection{Signal set selection}
To estimate codewords $\bar{v}_1$ and $\bar{v}_2$ based on the noisy observed superposition coded sequence $\bar{y}$, a partition of a signal constellation ${\cal M}$ into two parts is required. For QPSK modulation, a good partition into a Sidon set was found to be ${\cal M} = {\cal M}_1 \cup {\cal M}_2$, with
\[ {\cal M}_1 = \{-{1\over \sqrt{2}}, {1\over \sqrt{2}}\}, \quad {\cal M}_2 = \{-{j\over \sqrt{2}}, {j\over \sqrt{2}}\}, \]
where $j=\sqrt{-1}$ represents the quadrature component. In the absence of noise, the four received signal points in the direct-sum ${\cal M}_1+{\cal M}_2$ form a symmetric QPSK constellation. In a general case, an $N$-dimensional constellation needs to be partitioned as a {\em Sidon set}~\cite{uep_ldpc}. One way to achieve two levels of error protection for broadcasting over AWGN or Rayleigh fading channels is to select sets of signal points ${\cal M_i}$, $i-1,2$, such that two different values of minimum Euclidean distance (MED) $s_2>s_1$ are obtained, where
\begin{eqnarray*}
s_1 & = & \min_{\bar{v}_1^\prime\ne \bar{v}_1} \{ m_1(\vert \bar{v}_1\vert \bar{v}_1\vert) + m_2(\vert \bar{0} \vert \bar{v}_2\vert), \cr
& & \quad\qquad m_1(\vert \bar{v}_1^\prime\vert \bar{v}_1^\prime\vert) + m_2(\vert \bar{0} \vert \bar{v}_2\vert) \}\cr
&&\cr
s_2 & = & \min_{\bar{v}_2^\prime\ne \bar{v}_2} \{ m_1(\vert \bar{v}_1\vert \bar{v}_1\vert) + m_2(\vert \bar{0} \vert \bar{v}_2\vert), \cr
& & \quad\qquad m_1(\vert \bar{v}_1\vert \bar{v}_1\vert) + m_2(\vert \bar{0} \vert \bar{v}_2^\prime\vert) \}.\cr
\end{eqnarray*}

\vspace{-1em}
Note also that, by virtue of the $\vert u\vert u+v\vert$ construction, two values of Hamming (symbol) distances are obtained and thus two values of product distances are achieved. This means that with transmission over Rayleigh fading channels, two levels of diversity order are obtained.

With the QPSK constellation partition described above, the MED values are proportional to the square roots of the minimum Hamming distances of the constituent codes $C_1$ and $C_2$. That is, $s_i = \sqrt{2d_i}$, $i=1,2$. Therefore, with a QPSK signal constellation properly partitioned, binary LUEP codes can be applied to obtain a good superposition codes.

\section{Superposition coding and transmit diversity}
\label{sec3}
In the case of broadcasting information over wireless channels, and assuming block-fading Rayleigh statistics, superposition coding can be readily combined with the transmit diversity scheme introduced by Alamouti~\cite{alamouti}. This results in doubling the diversity order of superposition coding based on the $\vert u\vert u+c\vert$ construction with a relatively small impact on receiver complexity. 

It is remarked that the application of transmit diversity to cooperative wireless networks was introduced in~\cite{[9]} where the broadcasting nature of the problem was neglected (i.e., only one level of error protection was provided). Also, another advantage of the use of two-branch transmit diversity is its robustness to channel estimation errors~\cite{kurt}. 

Alamouti's scheme can be applied to superposition coding after modulation at the two source nodes~\footnote{Each transmitting node here is treated as a transmit antenna element in~\cite{alamouti}.} such that the transmitted sequences are, respectively
\begin{eqnarray*}
& \left[ m_1(\vert \bar{v}_1(t)\vert \bar{v}_1(t)\vert), \; -m_2^*(\vert \bar{0} \vert \bar{v}_2(t+nT)\vert) \right]\cr
& \left[ m_2(\vert \bar{0} \vert \bar{v}_2(t)\vert),  \; \; m_1^*(\vert \bar{v}_1(t+nT)\vert \bar{v}_1(t+nT)\vert) \right]
\end{eqnarray*}
where for convenience, the time index $t$ (or transmission time) of a sequence $m_i(\bar{v})$ is expressed as $m_i(\bar{v}(t))$, $i=1,2$. At the receiver, component-wise maximum ratio combining (MRC) is performed just as indicated in~\cite{alamouti}. However, instead of producing estimates of the transmitted bits, the output of the MRC stage is delivered to a demapper prior to soft-decision decoding. 

Finally, as the proposed superposition coding scheme is the combination of Plotkin's binary LUEP code construction and Alamouti's transmit diversity scheme, it is referred to as {\em Plotkin-Alamouti} coding.

\section{A two-level Plotkin-Alamouti superposition code}
\label{sec4}
A specific construction of a two-level superposition code based on the $\vert u\vert u+v\vert$ construction is presented in the following. Let $C_1$ be a binary single-parity-check $(20,19,2)$ code and $C_2$ be a binary Gallager LDPC $(20,7,6)$ code~\cite{gallager}. Plotkin's construction results in a $(40,26)$ LUEP code $C$ with separation vector $\bar{s}=(6,4)$ for the message space $M = \{0,1\}^7\times\{0,1\}^{19}$. In other words, codewords associated with 7 bits (resp. 19) have a minimum Hamming distance equal to 7 (resp. 4). 

Remarkably, since both component binary linear codes have low-density parity-check matrices, belief-propagation decoding can be applied based on the Tanner graph of $C$ to estimate codewords $\bar{v}_1$ and $\bar{v}_2$ based on the noisy received vector $\bar{y}$ described in section~\ref{sec2}. 

\begin{figure}[ht]
\centering
\includegraphics[width=3.4in]{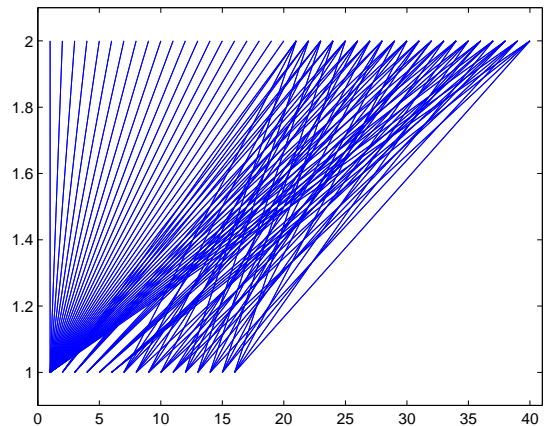}
\caption{The Tanner graph of a binary LUEP code based on an SPC (20,19,2) code and a Gallager LDPC (20,7,6) code.} 
\label{tanner}
\end{figure}

Fig.~\ref{tanner} shows the Tanner graph of LUEP code $C$ from which a sense of its UEP capabilities can be gained from direct examination. In particular, observe that those variable nodes in the second (right-most) part of a codeword clearly have higher degrees. These are nodes associated with information bits that can be recovered with high reliability by destination/relay nodes that are far away from the source nodes.
However, as noted in~\cite{uep_ldpc}, there is a ``vague association of UEP to different (node) degree distributions.'' 

\begin{figure}[ht]
\centering
\includegraphics[width=3.77in]{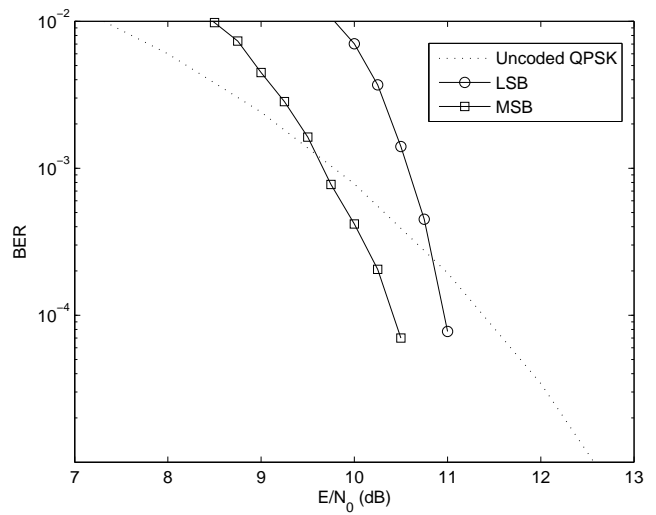}
\caption{Simulation results of a two-level superposition code for cooperative broadcasting over an AWGN channel with iterative BP decoding and a maximum of 50 iterations.} 
\label{simulation_awgn}
\end{figure}

Despite a large number of four-cycles, error performance is relatively good as shown in Fig.~\ref{simulation_awgn}, reporting the simu\-\-lated error performance of the superposition code based on $C$ for two cooperative source nodes broadcasting over AWGN channels with the QPSK constellation partition described in Section~\ref{sec2}. 

To obtain the simulation results, it was assumed that the decoder has knowledge of the AWGN power. A demapping stage takes place prior to decoding to obtain bit likelihood metrics, much in the same was as in bit-interleaved coded modulation~\cite{tarico}. Simulation results were obtained via iterative BP decoding with a maximum of 50 iterations. Transmission of 10,000 sequences was simulated. Moreover, the decoding process was stopped early upon finding a valid codeword. 

\vspace{1em}

(NOTE: The simulation results reported here are still preliminary. Completed and verified  simulation results for two-node cooperative broadcasting over both AWGN and Rayleigh fading, with Alamouti transmit diversity and MRC, will be available in March for the final version of the paper.)

\vspace{1em}

\section{Final remarks}
\label{sec5}
A construction technique of superposition codes for cooperative broadcasting over AWGN and Rayleigh fading channels has been introduced.  Two modulated subcode sequences are transmitted by two cooperating source nodes. The antennas at a destination/relay node effectively perform ``over-the-air mixing'' of these sequences, as suggested by Bergmans and Cover, and achieve unequal-error-protection capabilities: A receiving node close to the source node is able to estimate all the information bits with a high degree of reliability. Those nodes that are far away from the source pair are still able to recover a portion of the information bits. Extensions to other signal constellations and component error correcting codes is under way. The paper has also served to draw attention to the fact that cooperative broadcasting was introduced and analyzed thirty five years ago.

For  cooperative broadcasting over wireless networks that utilize block-fading Rayleigh channel models, the Alamouti transmit diversity scheme can be applied to both increase the diversity order and improve the robustness to channel estimation errors. Finally, the problem of finding good partitions of high-order signal constellations that form Sidon sets is a challenging one worth of further consideration.

\vspace{1em}

\section*{Acknowledgment}
The author gratefully acknowledges partial support of this work by the Japanese Society for the Promotion of Science (JSPS) under grant S-08185.

\vspace{1em}




%

\end{document}